\documentclass[twocolumn,twocolappendix]{aastex631}
\usepackage{multirow}
\pdfoutput=1

\newcommand{\target}{V\,1298 Tau}
\newcommand{\xmm}{{XMM-Newton}}

\newcommand{\fxu}{{erg s$^{-1}$ cm$^{-2}$}}

\newcommand{\lxu}{{erg s$^{-1}$}}

\received{November 18, 2021}
\accepted{December 3, 2021}

\shorttitle{Evaporation of V1298\,Tau exoplanets}
\shortauthors{A. Maggio et al.}
\graphicspath{{./}{figures/}}
\begin{document}

\title{New constraints on the future evaporation of the young exoplanets in the V1298 Tau system
% \footnote{Based on observations made with the Italian {\it Telescopio Nazionale Galileo} (TNG) operated by the {\it Fundaci\'on Galileo Galilei} (FGG) of the {\it Istituto Nazionale di Astrofisica} (INAF) at the {\it  Observatorio del Roque de los Muchachos} (La Palma, Canary Islands, Spain), under GAPS2 programme.}
}

\author[0000-0001-5154-6108]{A. Maggio}
\affiliation{INAF -- Osservatorio Astronomico di Palermo, Piazza del Parlamento, 1, I-90134, Palermo, Italy}

\author[0000-0002-9824-2336]{D. Locci}
\affiliation{INAF -- Osservatorio Astronomico di Palermo, Piazza del Parlamento, 1, I-90134, Palermo, Italy}

% \collaboration{6}{(GAPS Collaboration)}

\author[0000-0003-4948-6550]{I. Pillitteri}
\affiliation{INAF -- Osservatorio Astronomico di Palermo, Piazza del Parlamento, 1, I-90134, Palermo, Italy}

\author[0000-0002-4638-3495]{S. Benatti}
\affiliation{INAF -- Osservatorio Astronomico di Palermo, Piazza del Parlamento, 1, I-90134, Palermo, Italy}

\author[0000-0001-7707-5105]{R. Claudi}
\affiliation{INAF -- Osservatorio Astronomico di Padova, Vicolo dell'Osservatorio 5, I-35122, Padova, Italy}

\author[0000-0001-8613-2589]{S. Desidera}
\affiliation{INAF -- Osservatorio Astronomico di Padova, Vicolo dell'Osservatorio 5, I-35122, Padova, Italy}

\author[0000-0002-9900-4751]{G. Micela}
\affiliation{INAF -- Osservatorio Astronomico di Palermo, Piazza del Parlamento, 1, I-90134, Palermo, Italy}

\author[0000-0001-9984-4278]{M. Damasso}
\affiliation{INAF -- Osservatorio Astrofisico di Torino, Via Osservatorio 20, I-10025, Pino Torinese (TO), Italy}

\author[0000-0002-7504-365X]{A. Sozzetti}
\affiliation{INAF -- Osservatorio Astrofisico di Torino, Via Osservatorio 20, I-10025, Pino Torinese (TO), Italy}

\author[0000-0002-3814-5323]{A. Suarez Mascare{\~n}o}
\affiliation{Instituto de Astrofisica de Canarias, E-38205 La Laguna, Tenerife, Spain}
\affiliation{Departamento de Astrofisica, Universidad de La Laguna, E-38206 La Laguna, Tenerife, Spain.}



\begin{abstract}
Transiting planets at young ages are key targets for improving our understanding of the evolution of exo-atmospheres. We present results of a new X-ray observation of \target\ with XMM-Newton, aimed to determine more accurately the high-energy irradiation of the four planets orbiting this pre-main-sequence star, and the possible variability due to magnetic activity on short and long time scales. Following the first measurements of planetary masses in the \target\ system, we revise early guesses of the current escape rates from the planetary atmospheres, employing our updated atmospheric evaporation models to predict the future evolution of the system. Contrary to previous expectations, we find that the two outer Jupiter-sized planets will not be affected by any evaporation on Gyr time scales, and the same occurs for the two smaller inner planets, unless their true masses are lower than $\sim 40$\,$M_\oplus$. These results confirm that relatively massive planets can reach their final position in the mass-radius diagram very early in their evolutionary history.
\end{abstract}

\keywords{exoplanet systems --- exoplanet atmospheres --- X-rays stars --- pre-main sequence stars}


\section{Introduction} \label{sec:intro}
Planets around young stars deserve special attention, because they allow us to explore the evolutionary history during the first tens of million years after planetary formation. Moreover, the discovery of planetary systems with different architecture with respect to our Solar System has challenged the models of formation and evolution of planets.
 
One of the first and most puzzling young systems discovered so far with Kepler (K2 mission) is named after the host star \target, a K1 pre-main-sequence dwarf with a near-solar mass, belonging to the Group 29 stellar association \citep{Oh+2017}. \citet{David+2019b} announced a multiple system, including two Neptune-sized planets (dubbed "c" and "d"), one Jovian planet ("b"), and one Saturn-sized planet ("e"), in order of distance from the central star. The three inner planets have well-constrained orbital periods of $\sim 8, 12$, and $24$ days, while the outer planet was detected with a single transit event, and hence the orbital period was much more uncertain (40--120 days).
 
The relatively large planet sizes and rough mass estimates based on simple dynamical arguments suggested as most appealing the scenario of "fluffy" planets still in the contraction phase, due to radiative cooling and/or ongoing photoevaporation processes \citep{David+2019b,Poppenhaeger+2020}.

In this respect, an assessment of the high-energy environment
of the planets is very important, because changes of their masses and sizes on Gyr time scales may depend on the effect of stellar X-rays and UV irradiation onto the primary planetary atmospheres. A wide range of possible evolutionary paths has been recently investigated by \citet{Poppenhaeger+2020}, but limited by ignorance of the planetary masses and by quite different assumptions on the time evolution of stellar activity.

Intriguing new results from a program of optical spectroscopic observations were presented by \citet{paperI+2021} (hereafter SM21). This follow-up campaign provided RV time series sufficiently long to measure the masses of the two outer planets, 
$M_{\rm p,b} = 203\pm60$\,$M_{\oplus}$ and $M_{\rm p,e} = 369\pm95$\,$M_{\oplus}$, and to put constraints on the masses of the two inner planets, $M_{\rm p,c} < 76.3$\,$M_{\oplus}$ and $M_{\rm p,d} < 98.5$\,$M_{\oplus}$. An orbital period of $40\pm1$\,d was also determined for planet e. 
All the planetary masses (or upper limits), resulted significantly larger than guessed by \citet{David+2019b}, and even more so with respect to the lower expectations in the "fluffy-planet" scenario, where radii are presumed considerably larger at young ages compared to older planets with the same mass.

In this paper, we discuss first the expected evolutionary path of \target\ in the H-R diagram (Sect. \ref{app:A}). Next, we present new \xmm\ observations of the host star (Sect. \ref{sec:xuv}), aimed to determine the possible variability of the high-energy irradiation on short time scales and the spectral hardness of the stellar emission. Then, we revise the most likely evolution of the stellar activity (Sect. \ref{app:C}), and we explore how different models of the X/EUV flux ratio influence the forecasts on the fate of the system, employing a numerical modeling approach developed by \citet{Locci+2019} and further refined for the present work (Sect. \ref{sec:models}).
We discuss the results and draw conclusions in Section \ref{sec:discuss}.

% Figure 1 was here in v2
% \input{tabxspec1} old
% \input{deluxtabxspec} v2
% \vspace{-1.0cm}

\section{Star age and evolutionary path}
\label{app:A}
\target\ is a K1 star with a mass of $1.17\pm0.060$\,$M_{\odot}$, a radius of $1.278\pm0.070$\,$R_{\odot}$, an effective temperature $T_{\rm eff} = 5050\pm100$\,K, and a bolometric luminosity $L_{\rm bol} = 0.954\pm0.040$\,$L_{\odot}$ (SM21). It is located at a distance of $108.6\pm0.7$\,pc toward the Taurus region, and it belongs to the Group 29 stellar association \citep{Oh+2017}.

\begin{figure}[!t]
\begin{center}
\resizebox{\columnwidth}{!}{ \includegraphics[width=\hsize, bb=30 378
478 712]{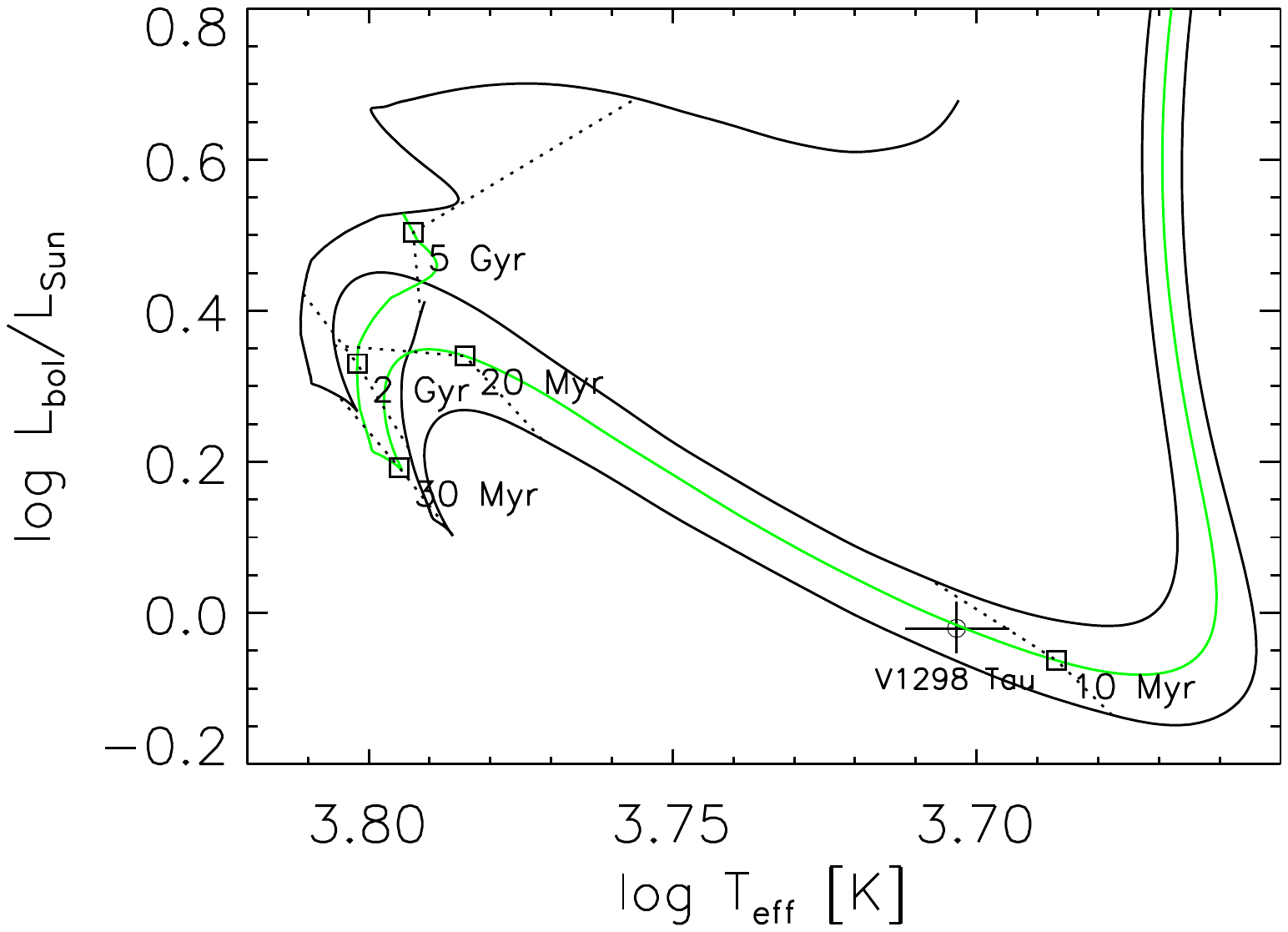} }
\end{center}
 \caption{Evolutionary track for \target\ (in green) in the temperature--luminosity diagram, according to the PISA models (\citealt{PISA2012}). Adjacent tracks (in black) are for stars of 1.1 $M_\odot$ and 1.2 $M_\odot$. Dotted segments join points at selected ages indicated near the square symbols.
 \label{fig:evoltrack} }
\end{figure}

Given the young age of our planet-hosting star, it is important to assess its current position in the theoretical temperature--luminosity diagram, and its evolutionary path up to the age of mature solar-type stars like the Sun. To this aim, we have adopted the PISA evolutionary tracks \citep{PISA2012}. In Figure \ref{fig:evoltrack} we show the current position of \target\ with respect to the tracks for stars of 1.1\,$M_\odot$ and 1.2\,$M_\odot$. We also show an intermediate interpolated track that yields a mass of $\sim 1.15$\,$M_\odot$ for \target, in agreement with the value reported by SM21, and an age of $11.9^{+2.0}_{-3.2}$\,Myr, considering the uncertainties on the effective temperature and bolometric luminosity. 
This age falls in the range 10--30\,Myr reported by SM21 for members of the Group 29 stellar association \citep{Oh+2017}, but we note that this range is clearly too large to assess the future evolution of our target in the temperature--luminosity diagram. 
The position of \target\ with respect to the PARSEC isochrones (see Fig.\ 4 in SM21) suggests an age at the younger extreme of this range. In fact, using a Bayesian inference method, SM21 have also constrained the age of \target\ to $9\pm2$\,Myr, which is more in line with our current estimate.

The above analysis indicates that the bolometric luminosity of \target\ is just above the minimum reached at the end of the Hayashi phase. It will increase by about a factor of two in the next 10\,Myr, before landing on the ZAMS as a G-type star at an age of $\sim 30$\,Myr, and by more than a factor of three until an age of 5\,Gyr. 
This evolution in effective temperature and bolometric luminosity is relevant for assessing the change of the equilibrium temperature of the planets and hence the Jeans escape parameter of their atmospheres (Appendix \ref{app:D}).

% \begin{figure}[!t]
% \begin{center}
% \resizebox{\columnwidth}{!}{ 
% \includegraphics[width=\hsize]{rgb.png} }
% \end{center}
% \caption{\xmm\ EPIC RGB image of \target. Red: $0.3-1.0$ keV, Green: $1.0-3.0$ keV, Blue: $3.0-8.0$ keV. V1298 Tau and the two bright nearby stars are labelled. The regions for the extraction of events of source and background are also displayed. \label{fig:x} }
% \end{figure}

%-----------------------------Figure Start------------------------------
% \begin{minipage}[b]{0.45\textwidth}
% \includegraphics[width=\textwidth]{TBD.png}
% \end{minipage}
   
\begin{figure}[thb]
\begin{center}
% \resizebox{!}{5.5cm}{
% \includegraphics{spec_epic_2apec.pdf}
% \includegraphics[width=0.5\hsize,bb=78 462 282 702]{all_3T_vapec_rgs_NHfixed.pdf}
\includegraphics[width=\hsize,bb=16 0 710 576]{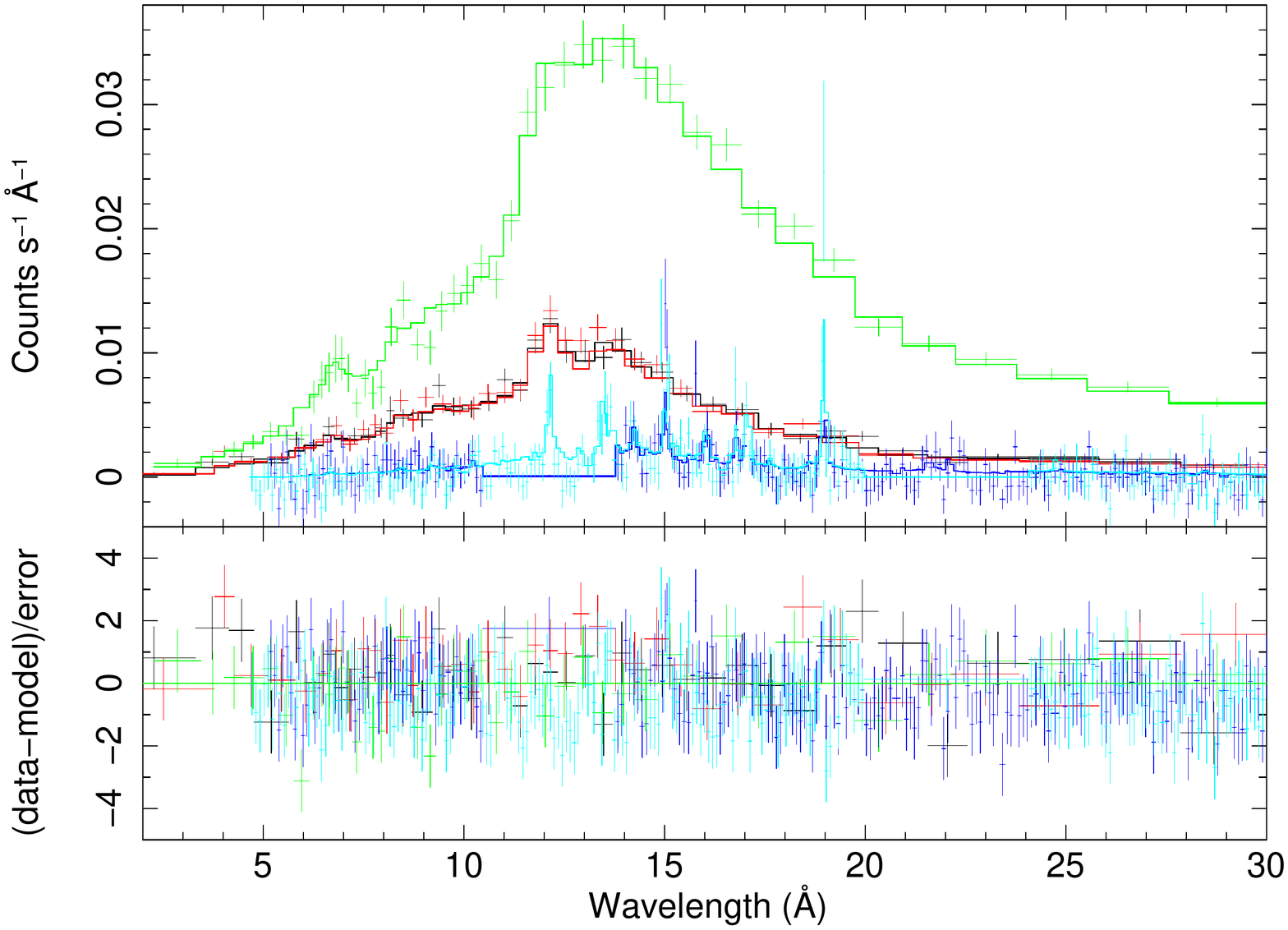}
\includegraphics[width=\hsize]{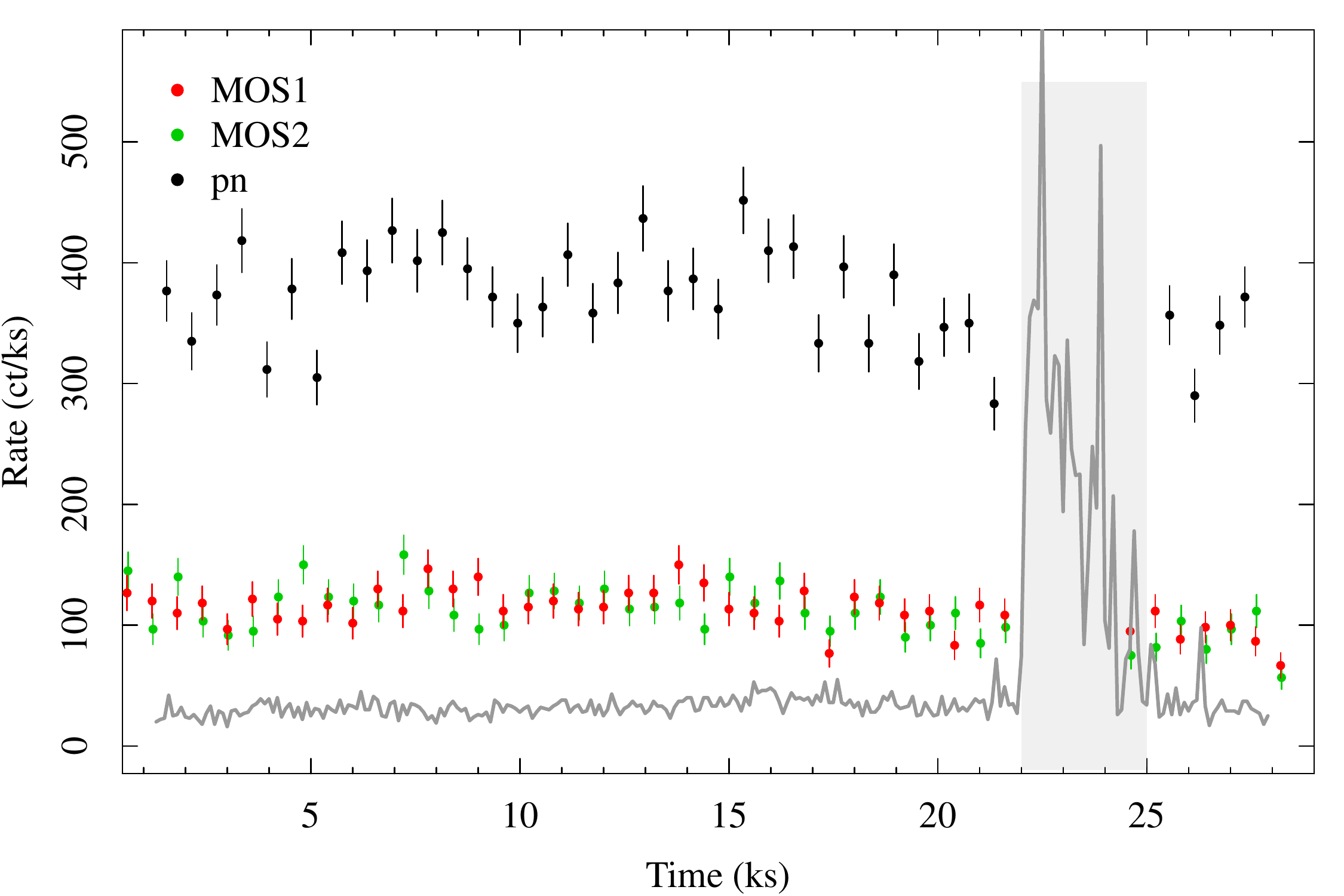} 
% }
\end{center}
\caption{\footnotesize{Upper: EPIC and RGS spectra of \target\ with best-fit model and residuals. Lower: pn and MOS light curves with bin size of 600\,s. The gray area marks the region of high background (gray curve), filtered out for the spectral analysis.  
\label{xraylc}  
}}
 \end{figure}
%-----------------------------Figure End--------------------------------

% \input{deluxtabxspec}

\begin{splitdeluxetable*}{cccccccccccBcccccccc}
% \tablenum{1}
\tablecaption{Best-fit parameters from modeling of the EPIC and RGS spectra of \target. 
\label{tabxspec}}
\tabletypesize{\scriptsize}
\tablewidth{0pt}
% \resizebox{\textwidth}{!}{
\tablehead{
\colhead{$N_\mathrm H$} & \colhead{T1} & \colhead{EM1} & \colhead{T2} & \colhead{EM2} & \colhead{T3} & \colhead{EM3} & \colhead{$\chi^2$} & \colhead{d.o.f.} &
% \colhead{Prob} & 
\colhead{$f_\mathrm X$} & 
\colhead{$L_\mathrm X$} & & \multicolumn{7}{c}{Abundances and 90\% confidence ranges (solar units; Anders \& Grevesse 1989)} \\
\colhead{$10^{20}$\,cm$^{-2}$} & \colhead{$10^6$\,K} & \colhead{$10^{52}$\,cm$^{-3}$} & \colhead{$10^6$\,K} & \colhead{$10^{52}$\,cm$^{-3}$} & \colhead{$10^6$ K} & \colhead{$10^{52}$\,cm$^{-3}$} & & & 
\colhead{$10^{-12}$\,erg s$^{-1}$ cm$^{-2}$} & \colhead{$10^{30}$\,erg$~$s$^{-1}$} & &
Mg               & Fe                  & Si                  & C                 & O            & N                        & Ne \\
& & & & & & & & & & & FIP (eV) &
7.65 & 7.90 & 8.15 & 11.26 & 13.62 & 14.53 & 21.56
}
% \decimalcolnumbers
\startdata
\\
$1.0^{+3.3}_{-1.0}$ & $3.0^{+1.4}_{-0.5}$ & $2.8^{+3.2}_{-1.3}$      & $8.3^{+0.5}_{-1.0}$ & $7.1^{+2.6}_{-1.9}$            & $16.3^{+2.8}_{-3.1}$ & $4.1^{+2.6}_{-1.1}$         & 3818.08   & 5719  & $1.20^{+0.18}_{-0.11}$ & $1.69^{+0.24}_{-0.15}$ &
% \multirow{2}{1.2cm}{free $N_{\rm H}$}
free $N_{\rm H}$ & 0.24 & 0.16 & 0.19 & 0.64 & 0.24 & 0.51 & 0.99 \\
& & & & & & & & & & & & $[0.09,0.42]$ & $[0.11,0.21]$ & $[0.09,0.31]$ & $[< 1.74]$ & $[0.14,0.35]$ & $[0.05,1.11]$ & $[0.60,1.35]$ \\
$1.6$ (fixed) & $3.0^{+0.9}_{-0.4}$ & $3.3^{+1.3}_{-1.1}$        
& $8.1^{+0.7}_{-0.7}$ & $6.9^{+2.8}_{-1.9}$            & $15.1^{+3.6}_{-1.7}$ & $4.8^{+1.7}_{-1.8}$         & 3818.13   & 5720  & $1.24^{+0.02}_{-0.04}$ & $1.75^{+0.03}_{-0.05}$ &
% \multirow{2}{1.2cm}{fixed $N_{\rm H}$}
fixed $N_{\rm H}$ & 0.21 & 0.15 & 0.19 & 0.71 & 0.21 & 0.45 & 0.89 \\
& & & & & & & & & & & & $[0.10,0.36]$ & $[0.11,0.20]$ & $[0.10,0.29]$ & $[< 1.71]$ & $[0.15,0.31]$ & $[0.05,1.00]$ & $[0.65,1.23]$ \\
\enddata
\tablecomments{Unabsorbed X-ray flux and luminosity in the 0.1--10 keV band. Errors are quoted at the 90\% confidence level.}
% }
\end{splitdeluxetable*}

\section{XMM-Newton observation}
\label{sec:xuv}
The \xmm\ observation of \target\ was performed on 2021 February 24
(ObsId 0864340301, PI A.\,Maggio), with the aim
to assess the spectrum and time variability of the host star in X-rays, and hence to characterize the coronal activity level and the dose of high-energy radiation received by its young planets. 
The exposure time was about 30\,ks and the prime instrument was EPIC with {\sc full frame} window imaging mode and the {\sc Medium} filter.
Data were obtained with all the CCD-based EPIC cameras and also with the high-resolution RGS spectrometers.

The Observation Data Files were reduced with the Science Analysis System (SAS, ver.18.0.0), following standard procedures. 
We obtained FITS lists of X-ray events detected with all the EPIC CCD cameras (MOS1, MOS2, and pn) and with the two high-resolution spectrographs (RGS1 and RGS2), calibrated in energy, arrival time, and astrometry with the {\sc evselect} SAS task. 
Inspection of the light curve of events detected with energies $> 10$\,keV, allowed us to identify and filter out a relatively small time interval affected by high background.

% \input{tabxspec2}
% Figure \ref{fig:x} shows the combined MOS+pn image of \target, with the nearby stars.There are 
We found
about 60 detected X-ray sources in the EPIC field of view, including two other bright young stellar objects: one is listed as a candidate YSO in SIMBAD (Gaia DR2 51884824140206720) and the other star is HD 284154, which appears also active and presumably young. The three sources have similar X-ray brightness, but HD 282154 appears redder because its spectrum is softer than the one of \target\ and the other YSO. 

\target\ is sufficiently isolated to allow the extraction of the source signal from
a circular region of 40\arcsec radii for MOS and pn, and local background from an uncontaminated nearby circular region of similar size. With SAS we also produced the response matrices
and effective area files needed for the subsequent spectral analysis. RGS source and background spectra were also extracted adopting the standard results of the SAS pipeline. Source X-ray spectra and light curves are shown in Fig. \ref{xraylc}.

% The quality of the data allowed us to perform a detailed spectral fitting which provides constraints on at least three thermal components and six element abundances.

For the spectral analysis, performed with {\sc xspec} V 12.10.1f, initially we applied a best-fitting procedure only to the EPIC (MOS1, MOS2, and pn) X-ray spectra. We adopted an optically thin coronal emission model composed by three isothermal components (3T, {\sc vapec}), with the abundances of all elements linked to the iron abundance. Eventually, we added also the RGS1 and RGS2 high-resolution spectra, and allowed up to six elements as free parameters: C, N, O, Ne, Mg, Si, and Fe (Table \ref{tabxspec}).
The source spectrum was also multiplied by a global interstellar absorption component ({\sc phabs}). A best fit was achieved with a reduced $\chi^2 = 0.67$ for 5719 d.o.f.
%, and probability P$(\chi^2>\chi^2_0) \sim 1$).

Elemental abundances follow a trend typical of young active stars \citep{Maggio+2007,Scelsi+2007}, i.e.\ elements with low First Ionization Potential (FIP), such as Fe, are systematically underabundant with respect to high-FIP elements, such as Ne, with the notable exception of oxygen. Since coronal Ne traces the actual photospheric abundance more closely than other elements \citep{Maggio+2007}, our measurement confirms a near-solar metallicity for \target\ (SM21).

However, this result should be taken with caution, because the \target\ coronal abundances are quantified with respect to the solar photospheric abundances \citep{AG1989}, while stellar abundances for each element should be employed for a proper assessment of any FIP effect \citep{SanzForcada+2004}.  

Since both the hydrogen column density, $N_{\rm H}$, and the C abundance were poorly constrained, we performed a final fit by fixing $N_{\rm H} = 1.6\times10^{20}$\,cm$^{-2}$, the value derived from the known B--V color excess, $E(B-V) = 0.024 \pm 0.015$ (\citealt{David+2019b}),
which implies an extinction $A_{\rm V} = 0.074 \pm 0.05$ and $N_{\rm H}$ in the range $6.2 \times 10^{19}$--$2.7 \times 10^{20}$\,cm$^{-2}$. We checked that this uncertainty implies just a 6\% additional systematic error on the measured X-ray flux.

The unabsorbed flux and the luminosity of \target\ are $f_{\rm x} = 1.19\times10^{-12}$\,\fxu\ and $L_{\rm x} = 1.68^{+0.03}_{-0.06}\times10^{30}$\,\lxu, respectively, in the band 0.1--2.4\,keV. The X-ray to bolometric luminosity ratio resulted $\log L_{\rm x}/L_{\rm bol} = -3.35^{+0.01}_{-0.02}$. These
values confirm that \target\ is an X-ray bright young star near the saturated emission regime
observed for G-K stars of similar age \citep{Pizz03,2012MNRAS.422.2024J}.

The X-ray light curves of \target\ (Fig. \ref{xraylc})
% (Fig. \ref{xraylc}, bottom panel) 
show time variability on order of
30\% with respect to the average rate level during the observation. No large flares were observed.

\citet{Poppenhaeger+2020} reported results based on a snapshot X-ray observation ($t_{\rm exp} \sim 1$\,ks) with Chandra, and spectral fitting assuming an isothermal model with fixed solar abundances and negligible interstellar absorption. For comparison, our results show a broad range of coronal plasma components with temperatures ranging from $\sim 3$ to $\sim 15$\,MK, and a pattern of chemical abundances typical of very active stars (see above).     
Our total measured X-ray flux
is about 25\% higher than the flux reported by \citet{Poppenhaeger+2020}. Considering also the early detection of \target\ in the ROSAT All Sky Survey, our
results imply very little variability of the quiescent X-ray emission on a time scale of about 30 yr, and comparable to the short-term variability that we observed in the time frame of the XMM-Newton observation ($\Delta t \sim 0.35$\,day).

\section{Evolution of XUV IRRADIATION}\label{app:C}
Several formulations were proposed in the past to describe how both the X-ray and the EUV irradiation change in time on Gyr time scales. For the present work, we tested two different descriptions:
the simple X-ray luminosity vs.\ age analytical relation proposed by \citet{Penz08a} (hereafter PM08) for G-type stars, and the most recent semi-empirical modeling by \citet{Johnstone+2021} (hereafter J21).

\begin{figure}[thb]
\begin{center}
% \resizebox{!}{5.5cm}{
\includegraphics[width=\hsize, bb=36 382 472 708]{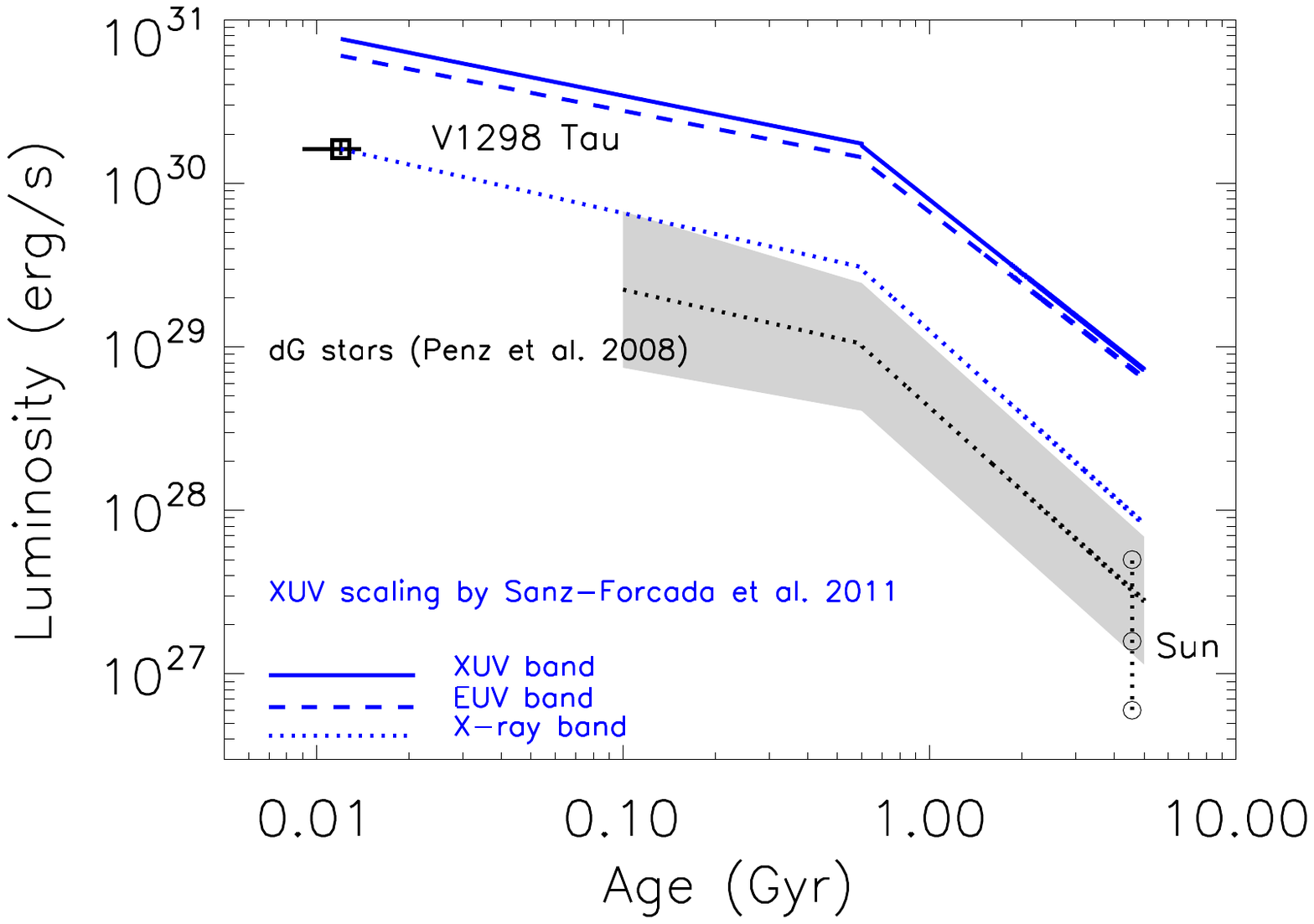}
\includegraphics[width=\hsize, bb=36 382 472 708]{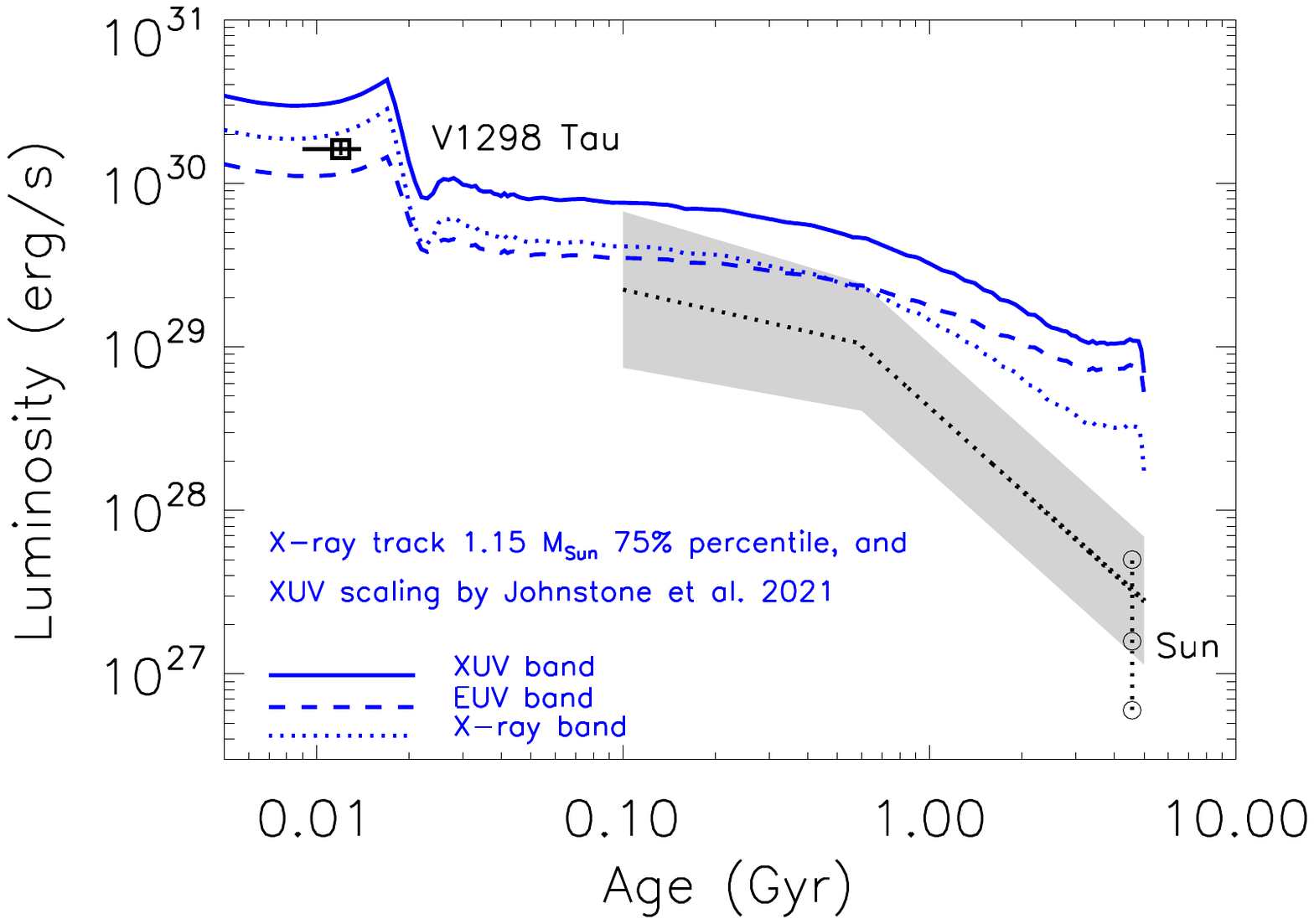} 
% }
\end{center}
\caption{\footnotesize{Upper: Time evolution of X-ray, EUV, and total XUV luminosity of \target, according to PM08 and the X-ray/EUV scaling by \citet{SF11}. The gray area is the original locus for dG stars in PM08. Lower: As above but adopting the evolutionary tracks of J21 for a 1.15\,$M_\odot$ star with rotation rate matching the measured value for \target.
\label{fig:xuvevol}  
}}
 \end{figure}

According to PM08, the coronal X-ray luminosity for G-type stars declines with age following a broken power law. It was calibrated considering the X-ray luminosity distributions of stars in the Pleiades and Hyades open clusters, together with field stars and the Sun. The evolutionary path is represented in Fig.\ \ref{fig:xuvevol}. The gray area indicates the observed $1\sigma$ spread around the median of the X-ray luminosity distribution for the two open clusters. We anchored the starting X-ray luminosity (5--100\,\AA)  of \target\ to the measured value (Sect.\ \ref{sec:xuv}), and predicted the future evolution following the same median power-law slopes. In order to predict the EUV luminosity (100--920\,\AA), we employed the scaling law proposed by \citet{SF11} (SF11).
% In the P08 case, we imposed the observed \lx\ at the current stellar age (Sect. \ref{sec:xuv}) as initial value.

Next, we adopted the description by J21, where the X-ray to bolometric luminosity ratio, $L_{\rm x}/L_{\rm bol}$, follows a broken power-law with the Rossby number, i.e.\ the ratio of the rotation period to the convective turnover time; in this scenario, the evolution of the stellar rotation plays a crucial role,
depending on the initial rotation rate at early ages ($\Omega_0$ at $\sim 1$\,Myr), the exchange of angular momentum between the stellar core and the convective envelope, the magnetic coupling between the star and the circumstellar disk, and the magnetic braking due to a magnetized stellar wind. All these effects are modeled and calibrated by comparison with observed distributions of stellar rotation periods for stars of different masses and ages. Although this model is clearly more sophisticated than the purely empirical law of PM08, we note that its predictions for a solar-mass star overestimate the observed median X-ray luminosity of the Sun by a factor $\sim 5$.

We selected as the most appropriate for \target\ the evolutionary path (Fig.\ \ref{fig:xuvevol}) for a 1.15\,$M_\odot$ star, with initial rotation rate $\Omega_0$ equal to the current measured value $\Omega = 9.36\pm0.16$\,$\Omega_{\odot}$\footnote{Assuming $\Omega_{\odot} = 2.67\times10^{-6}$\,rad\,s$^{-1}$.},
computed from the measured rotation period, $P_{\rm rot}=2.91\pm0.05$\,day (SM21).
This value of $\Omega_0$ corresponds to the 75\% percentile of the observed distribution of rotation rates considered by J21 for stars with similar mass. Our choice is justified self-consistently by the prediction that a star with the mass, age, and rotation rate of \target\ maintains an almost constant value of $\Omega$ from 1\,Myr to few hundreds of million of years. In fact, this educated guess of the initial rotation rate
identifies an evolutionary track of stellar activity that predicts quite precisely the observed X-ray luminosity of \target and hence defines its most probable future evolution.

Following J21, we have also retrieved the evolutionary path of the EUV luminosity (100--920\,\AA), predicted on the basis of an empirical mass-independent power-law scaling between the surface EUV and X-ray fluxes, calibrated on a sample of late-type stars observed with the EUVE satellite and solar spectra derived from the TIMED/SEE mission.
% Recall: J21 X-ray range is 5.17-124 \AA (overlap with EUV band implies overestimation of total XUV irradiation by few percent (J21, Sect. 4.1).

% Both describe the evolution of the X-ray luminosity, % L$_{\rm x}$, with a saturation and a decay phase. 
% We note that at each age stars show a spread in X-ray luminosity that is associated with the spread of the stellar rotation rates (\citealt{Pizz03}. As a consequence, stars that are fast rotators at young age, namely after disc dissipation, remain in the saturation regime for longer times  (\citealt{Tu15}), leading to different levels of XUV radiation at which planets could be subjected during their early evolutionary stages. 
Figure \ref{fig:xuvevol} clearly shows that the main difference between the PM08+SF11 and J21 descriptions rests on the X-ray/EUV scaling rather than on the X-ray luminosity evolution. In particular, J21 predict a significantly lower EUV luminosity, which becomes larger than the X-ray luminosity only for ages $\ga 600$\,Myr. Instead,
the SF11 scaling yields higher EUV irradiation at any age, but with a faster decline over time.

To understand the effects of these different descriptions in our photoevaporation modeling, we have also considered a hybrid case in which the X-ray luminosity evolves according to J21 but the EUV luminosity is computed with the SF11 scaling.
In summary, we explored three different parameterizations: the analytical formulation by PM08 for the X-ray evolution coupled with the SF11 scaling for the EUV band (PMSF model), the full numerical description by J21 (JoJo model), and a hybrid case with the J21 X-ray evolution coupled with the SF11 scaling (JoSF model).

\begin{table*}[t!]
\centering
\caption{Modeling results of atmospheric photoevaporation for planets c and d.} 
% \scriptsize
\begin{tabular}{ccccccccccccccccc}
\hline\hline  
\multicolumn{5}{c}{ }  & &
\multicolumn{3}{c}{ PMSF}  &  & \multicolumn{3}{c}{ JoSF} & & \multicolumn{3}{c}{ JoJo}\\ 
\cline{7-9}
\cline{11-13}
\cline{15-17}

& $M_p$ & $R_p$ & $M_c$ &  $R_c$ & & $M_f$ &  $R_f$ & $\Delta$t & &  $M_f$ &  $R_f$ & $\Delta$t & &  $M_f$ &  $R_f$ & $\Delta$t \\
Case & (M$_\oplus$) & (R$_\oplus$) & (M$_\oplus$)& (R$_\oplus$) & & (M$_\oplus$) & (R$_\oplus$)& (Gyr) & & (M$_\oplus$) & (R$_\oplus$)& (Gyr) & & (M$_\oplus$) & (R$_\oplus$) & (Gyr) \\
\hline 
\multicolumn{17}{c}{Planet c} \\
${\rm M_1}$ & 38.1 & \multirow{4}{1.8cm}{$5.05\pm0.38$} & 37.1 & 2.64 & & 37.9 & 3.42 & 0.2$^a$ & & 38.0 & 3.44 & 0.22$^a$ & & 38.1 & 3.47 & 0.28$^a$ \\
${\rm M_2}$ & 25.4 & & 24.6 & 2.37 & & 24.8 & 2.71 & & & 24.8 & 2.72 & & & 25.1 & 3.05 & \\
${\rm M_3}$ & 20.6 & & 20 & 2.22 &  & 20 & 2.22 & 1.55$^b$ & & 20 & 2.22 & 2.7$^b$ & & 20.16 & 2.7 & \\
${\rm M_4}$ & 10.3 & & 10 & 1.84 & & 10 & 1.84 & 0.03$^b$ & & 10 & 1.84 & 0.03$^b$ & & 10 & 1.84 & 0.16$^b$ \\
\multicolumn{17}{c}{Planet d} \\
${\rm M_2}$ & 32.6 & \multirow{3}{1.8cm}{$6.30\pm0.45$} & 30.5 & 2.53 & & 32.1 & 3.64 & 1.77$^a$ & & 32.2 & 3.68 & 2$^a$ & & 32.5 & 3.8 & 2.94$^a$ \\
${\rm M_3}$ & 21.4 & & 20 & 2.22 & & 20.2 & 2.79 & & & 20.3 &2.82 & & & 20.6 & 2.25 & \\
${\rm M_4}$ & 10.6 & & 10 & 1.84 & & 10 & 1.84 & 0.09$^b$ & & 10 & 1.84 & 0.16$^b$ & & 10 & 1.84 & 1.16$^b$ \\
\hline
\end{tabular}
\flushleft
\footnotesize
NOTES -- $M_p$ and $R_p$ are the initial planet mass and radius, $M_c$ and $R_c$ are the assumed core mass and radius, $M_f$ and $R_f$ are the final planet mass and radius at 5\,Gyr. $^a$ Time employed by the planet to become stable against evaporation.
$^b$ Time employed by the planet  to lose completely its atmosphere.
\label{Tab:par}
\end{table*}

\begin{figure*}
\gridline{\fig{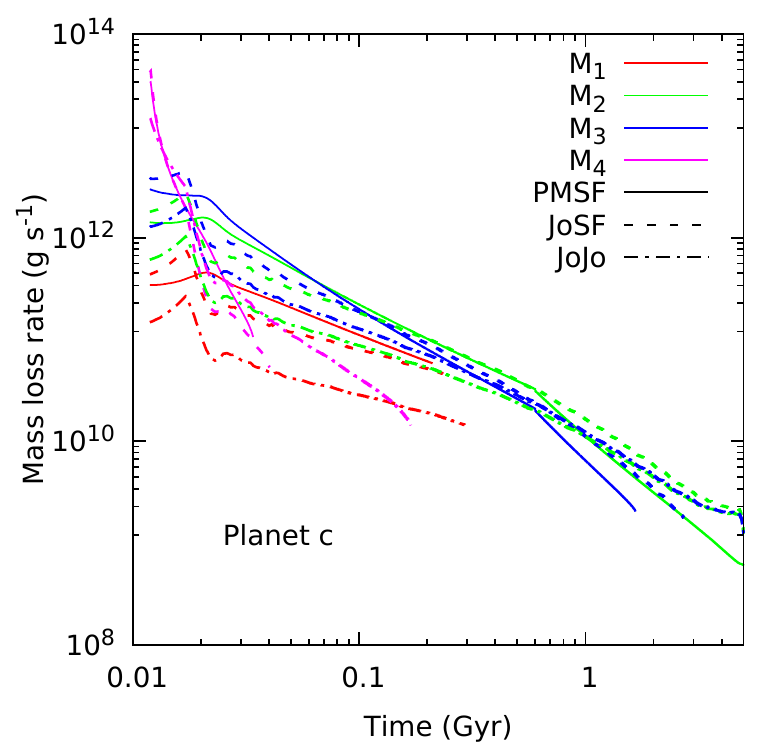}{0.33\textwidth}{}
          \fig{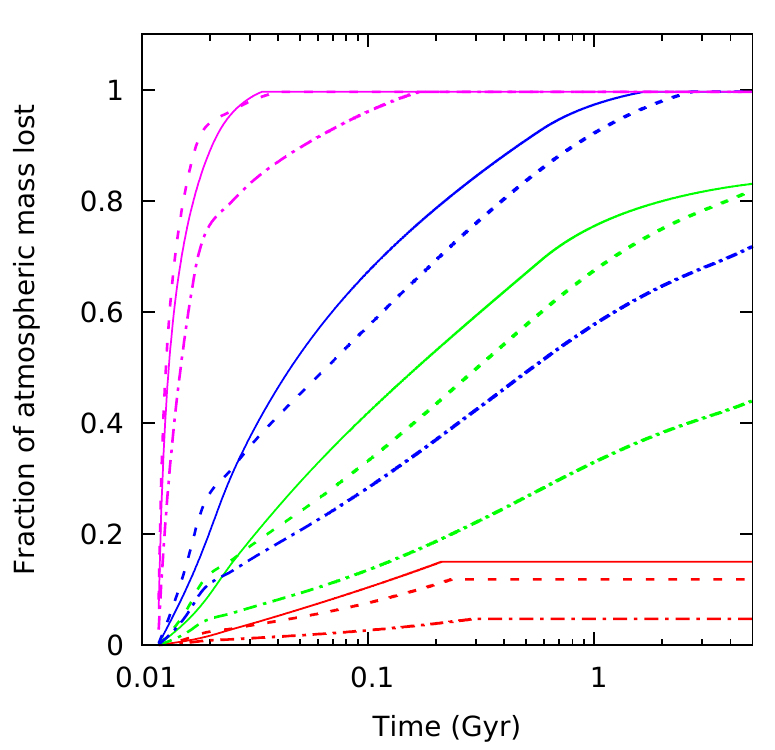}{0.33\textwidth}{}
          \fig{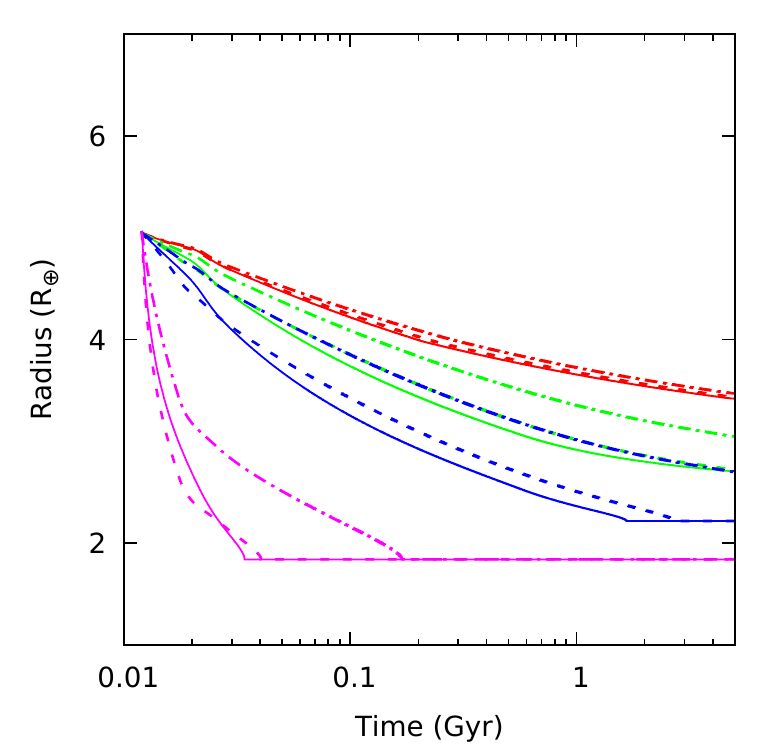}{0.33\textwidth}{}}
          \vspace{-1.0cm}
\gridline{\fig{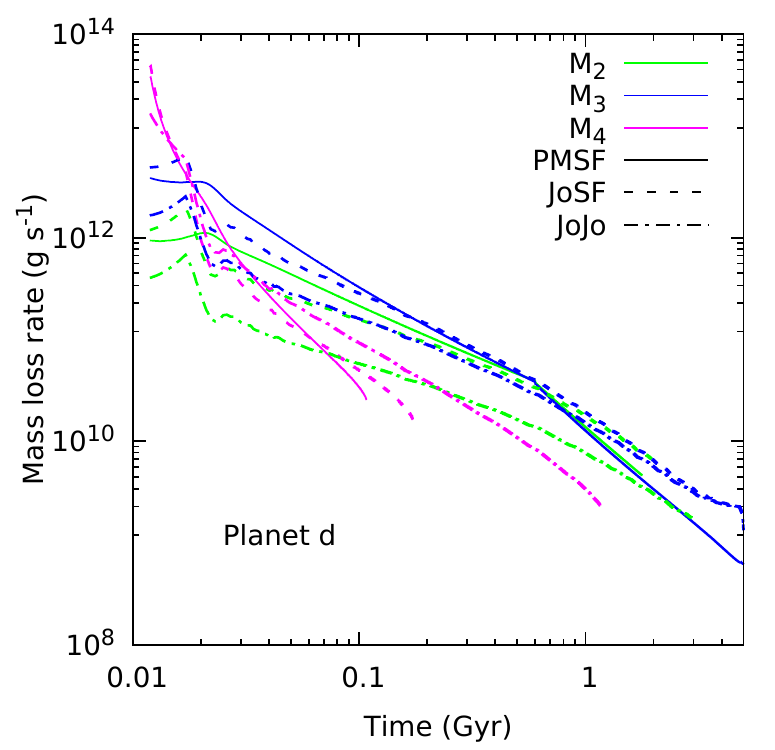}{0.33\textwidth}{}
          \fig{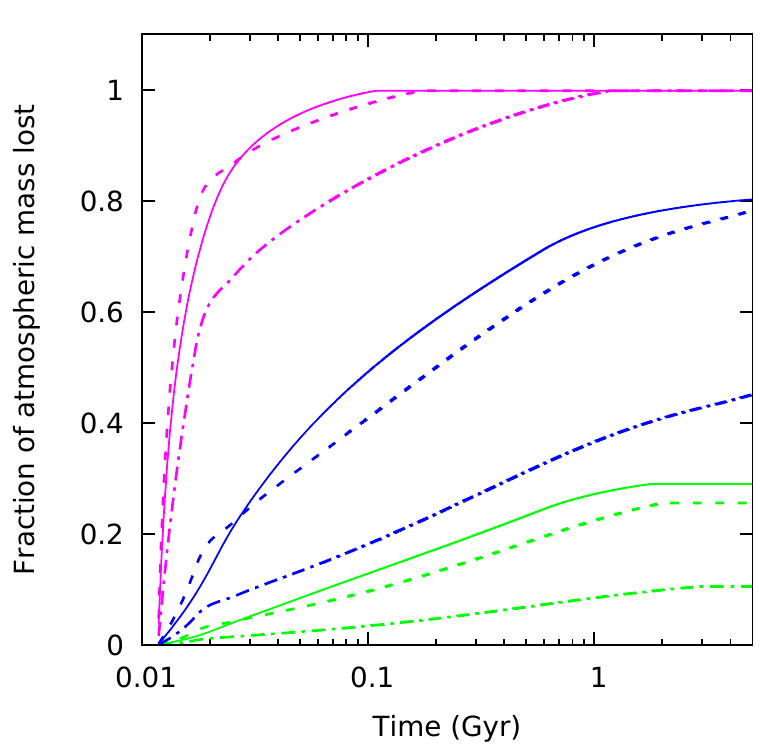}{0.33\textwidth}{}
          \fig{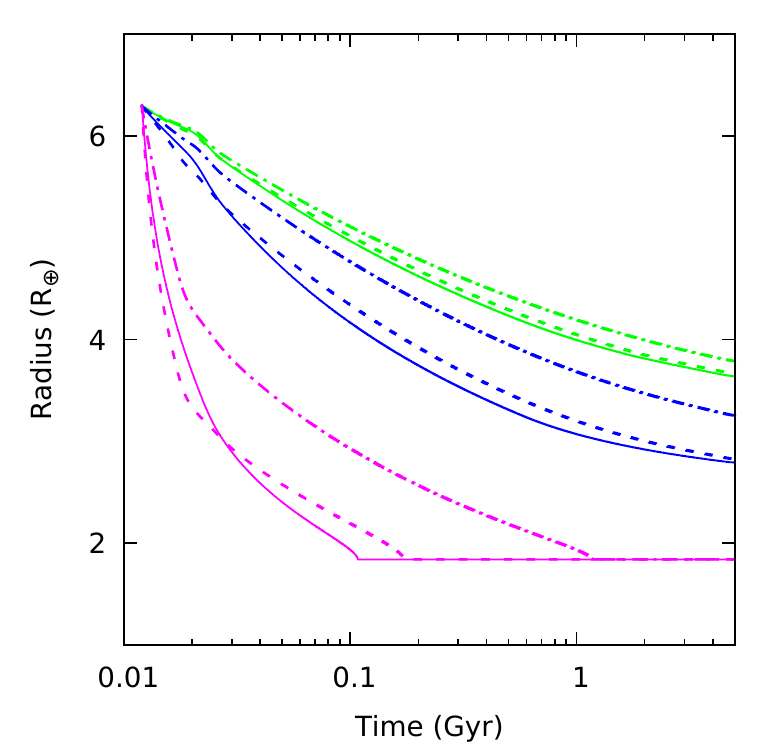}{0.33\textwidth}{}}
          \vspace{-0.5cm}
\caption{Plots of mass-loss rate, cumulative fraction of atmospheric mass lost, and planetary radius vs.\ time for planets c and d. In each panel, we use different colors and line styles for each assumed planetary mass and modeling approach.
% according to the different simulations explained in the text.
}   
\label{figure:MR_evolution}
\end{figure*}

\section{Planetary photoevaporation}\label{sec:models}
We investigated the predicted evolution of planetary atmospheres induced by the stellar high-energy radiation, taking into account the updated parameters reported in Sect. \ref{sec:intro}.

Proceeding similarly as in \citet{Georgieva2021} and in \citet{Benatti+2021}, we evaluated the mass-loss rate of the planetary atmosphere using the hydrodynamic-based approximation developed by \citet{kuby+2018a} for planetary masses lower than 40 $M_{\oplus}$, or the energy-limited approximation \citep{Erkaev+2007} for higher values.
Then we computed the change in atmospheric mass fraction, planetary mass, and radius, from the present age to 5\,Gyr, in steps of 1\,Myr, assuming a fixed circular planetary orbit (Appendix \ref{app:D}).
At any time step, we have employed the Jeans escape parameter \citep{Fossati+2017} to assess if a planet reaches stability against evaporation \citep{Kuby18b}.
A major improvement in our modeling approach with respect to previous applications consisted in taking into account the evolution of the bolometric luminosity (Sect. \ref{app:A}), which factors into the Jeans stability criterion and in determining the size of the gaseous envelope.

We explored different scenarios for the inner planets $c$ and $d$, having only upper limits on their masses. We considered five possible values: $M_0 = M_{\rm ul}$, i.e.\ the value of the upper limit, $M_1 = M_{\rm ul}/2$, $M_2 = M_{\rm ul}/3$; instead, $M_3$ and $M_4$ were computed assuming a value of 20 and 10\,$M_{\oplus}$ for the core mass, respectively, implying planetary masses between 10\% and 30\% the current upper limits (Table~\ref{Tab:par}). For comparison, the core mass of Saturn is loosely constrained in the range 5--20\,$M_\oplus$ \citep{RavitGuillot2013}.

As a matter of fact, we found that the most massive outer planets b and e are already stable against evaporation at the initial time. Hence, they are not expected to evolve appreciably. The same occurs for
planet c in the case $M_0$, and for planet d in both the cases $M_0$ and $M_1$. This is basically due to a combination of their high density and/or the relatively large distance from the host star. 

Instead, in all the other cases, 
% ($M_2$, $M_3$, $M_3$ and $M_5$ for planet c, and  $M_3$, $M_4$ and $M_5$  for  planet d), 
the planets c and d go through a hydrodynamical instability phase, whose intensity and duration depend on their mass. In Figure~\ref{figure:MR_evolution} and in Table~\ref{Tab:par} we report our results. The latter includes the core mass ($M_{\rm c}$) and radius ($R_{\rm c}$) for each case, and the planetary mass ($M_{\rm f}$) and radius ($R_{\rm f}$) at 5\,Gyr for each adopted XUV irradiation model. We also report the time ($\Delta t$) taken by the planet for losing completely its gaseous envelope or for reaching stability against evaporation, if these conditions occur. In the former case, the cumulative fraction of the atmospheric mass lost reaches 1 (Fig.\ \ref{figure:MR_evolution}, central panels) and the planetary radius remains equal to the core radius (same figure, right panels). Instead, the stability condition implies a residual non-null atmospheric mass fraction, but a radius that keeps slowly decreasing due to gravitational shrinking and stellar bolometric luminosity evolution. 

% In Figure~\ref{figure:MR_evolution} we show, for each case and model, the mass loss rate, the cumulative fraction of the atmospheric mass lost with respect to the initial value, and the planetary radius.

We found that stability can be reached only in the case $M_1$ for planet c, and case $M_2$ for planet d. For lower masses, the atmospheric mass-loss rate increases, in general, eventually leading to the complete disappearance of the atmosphere. This fate is met by both planets in the $M_4$ case, independently from the XUV irradiation model adopted. However, different models affect the time required to reach stability or to lose the atmosphere. The case $M_3$ for planet c is critical in this respect, because complete atmospheric evaporation occurs only if the \citet{SF11} X/EUV scaling is considered.

\section{Discussion} \label{sec:discuss}

Our analysis of the high-energy emission level of \target\ and its short-term variability suggest that this PMS star has a fairly high and steady activity level, typical of coronal sources in the saturated regime \citep{Pizz03}. Comparison of the observed X-ray luminosity with the predictions based on the semi-empirical models of rotation and activity evolution by \citet{Johnstone+2021} indicates that our target
followed a track with an almost constant rotation rate from 1\,Myr to the present age (9--14\,Myr). Following this working hypothesis, we can estimate that uncoupling of the star from the circumstellar disk occurred at an age $\tau_{\rm d} \approx 4.4$\,Myr \citep{Tu15}.
This is actually a lower limit to the disk lifetime, i.e.\ the time available for planets to migrate and accrete rocky material from the disk, and to consequently grow in mass and size.

The relatively large masses for the two outer planets, with respect to their radii, imply quite heavy cores and/or a high metallicity content (SM21),
that need to be explained by a detailed modeling of their migration histories. This occurrence implies that they are already stable against photoevaporation, and hence they should not move further from the present  position in the mass-radius diagram.

For the inner planets, 
% only upper limits on their masses are available. Taking into account the known radii, 
we conjecture that their characteristics are intermediate between those of Neptune and Saturn.
We considered a range of planetary masses 10--76\,$M_\oplus$ for planet c, and 10--98\,$M_\oplus$ for planet d, and estimated atmospheric mass fractions of $\sim 3$\% for c and 6\%--7\% for d. We found that they may be affected by photoevaporation only if their true masses are lower than about 40\,$M_\oplus$, and they will completely lose their atmospheres if their masses are lower than about 20\,$M_\oplus$.

In case of atmospheric photoevaporation, the end state of planetary evolution depends on several factors. The decrease in mass is relatively small with respect to the decrease in radius (by factors 1.5--3), hence the Jeans escape parameter (Eq.\ \ref{eq:J}) tends to increase and hydrodynamical stability could be reached. The opposite trend can be caused by an increase of the bolometric irradiation, leading to an increase of the planetary equilibrium temperature. Moreover, the atmospheric mass-loss rate is larger for planets with lower density, and hence planets with a lower initial mass may completely lose their gaseous envelope before stability is reached.

The mass-loss rate depends on the planetary characteristics that determine its gravitational potential and thermal energy of the atmosphere, but also on the time-variable high-energy irradiation. For planets c and d, our simulations with the models PMSF and JoSF provide similar results for any mass case. Hence, the simple analytical description of the X-ray luminosity evolution vs.\ age by PM08 results as much effective as the tabulated model by J21 when they are coupled to the same X-EUV scaling by SF11. On the other hand, the JoJo model yields a less vigorous photoevaporation, clearly due to the smaller EUV irradiation predicted at any age. This result shows that the choice of the scaling law for the computation of the total XUV irradiation is relevant in determining the time scale of atmospheric evolution, but the end fate of photoevaporation is identical in most cases.

On the other hand, if mass values near the current upper limits will be assessed for planets c and d with future measurements, their characteristics will not be affected by further atmospheric evolution. We conclude that current measurements are consistent with an evolution of all the four planets much faster than expected so far.

% {\bf commento di fare chiarezza su come evolve la luminosità EUV che ad oggi rimane un campo ancora non capito del tutto?}

\begin{acknowledgments}
We acknowledge financial contribution from the ASI-INAF agreement n.2018-16-HH.0 (THE StellaR PAth project), and from the ARIEL ASI-INAF agreement n.2021-5-HH.0. A.S.M.\ acknowledges financial support from the Spanish MICINN under 2018 Juan de la Cierva program IJC2018-035229-I.
We also acknowledge partial support by the projects PLATEA (ASI-INAF agreement n.2018-16-HH.0) and HOT-ATMOS (PRIN INAF 2019). A.M.\ thanks the anonymous referee and Dr. A.F. Lanza for useful comments on the manuscript.
Based on observations obtained with \xmm, an ESA science mission with instruments and contributions directly funded by ESA Member States and NASA.
\end{acknowledgments}


% \vspace{5mm}
\facilities{XMM (EPIC and RGS)}


\software{SAS
(\citealt{SAS2004}, \url{www.cosmos.esa.int/web/xmm-newton/sas}),
XSPEC (\citealt{XSPEC1996}, \url{heasarc.gsfc.nasa.gov/xanadu/xspec/})
% astropy \citep{2013A&A...558A..33A,2018AJ....156..123A},  
%          Cloudy \citep{2013RMxAA..49..137F}, 
%          Source Extractor \citep{1996A&AS..117..393B}
%          
}



% \newpage
\appendix
 
\section{Planetary evolution \label{app:D}}
According to \citet{Fossati+2017}, photoevaporation of planetary atmospheres occurs when the Jeans escape parameter
\begin{equation}
    \Lambda = \frac{G m_{\rm H} M_{\rm p}}{k_{\rm B} T_{\rm eq} R_{\rm p}} < 80
    \label{eq:J}
\end{equation}
where $G$ is the gravitational constant, $m_{\rm H}$ is the hydrogen mass, and $k_{\rm B}$ is the Boltzman constant, while $M_{\rm p}$ and $R_{\rm p}$ are the planet mass and radius, and $T_{\rm eq}$ is the planet equilibrium temperature, computed as
\begin{equation}
T_{\rm eq}=T_{\rm eff}\biggl[f_{\rm p}(1-A_{\rm B})\biggr]^{1/4}\biggl(\frac{R_{\rm *}}{2 d}\biggr)^{1/2}
\end{equation}
where $T_{\rm eff}$ and $R_{\rm *}$ are the stellar effective temperature and radius, respectively, $A_{\rm B}$ is the Bond albedo, $d$ the star--planet distance, and $f_{\rm p}$ is a parameter that takes into account whether the planet is tidally locked or not. We assumed $A_{\rm B} = 0.5$, as for Jupiter, and $f_{\rm p}=\frac{2}{3}$, because tidal locking is expected on time scales $\ll 1$\,Myr, for orbits with negligible eccentricity \citep{Leconte+2010,Ogilvie2014}. 
% $f_p=\frac{1}{4}$ not tidally locked

Through $T_{\rm eff}$, $\Lambda$ depends on the stellar bolometric luminosity, and both these parameters enter also in the analytic expression for the atmospheric mass-loss rate derived by \citet{kuby+2018a}, which is also a function of the XUV irradiation at the planetary orbital position.

% In addition we allow for the evolution of the stellar  effective temperature and bolometric luminosity. In fact the effective temperature of the star contributes to the determination of the Jeans escape parameter, whereas the bolometric luminosity  plays a role in the evaluation of the planetary radius.

For each initial value of $M_{\rm p}$, we calculated the core radius, $R_{\rm c}$, through the analytic relation proposed in \citet{Benatti+2021} and valid for planets with Earth-like composition. We then computed the radius of the atmospheric envelope as $R_{\rm env} = R_{\rm p} - R_{\rm c}$, and the initial atmospheric mass fraction, $f_{\rm atm} = M_{\rm env}/M_{\rm p}$, 
% We account for the evolution of the planetary radius 
by inverting the relation given in \citet{LopFor14}.
% already adopted, e.g., by~\citet{Poppenhaeger+2020} for the \target\ system. 
This relation was developed for atmospheres dominated by H--He, and provides $R_{\rm env}$ as a function of $M_{\rm p}$, the bolometric flux received, $f_{\rm atm}$, and the age of the system, taking into account also the gravitational shrinking. Furthermore, it allows the choice of solar or enhanced opacity of the atmospheric gas.

In turn, the atmospheric mass fraction allows the computation of the core mass, $M_{\rm c} = M_{\rm p} (1 - f_{\rm atm})$, which we assume constant in time.
Alternatively, if the core mass $M_{\rm c}$ is fixed {\it a priori}, we can estimate $M_{\rm p}$
iteratively, given $R_{\rm env}$ and $f_{\rm atm} = 1 - M_{\rm c}/M_{\rm p}$.
% In the case of Jovian or sub-Jovian planets for the evolution of the radius, we instead use the numerical models carried out by \citet{Fortney+2007} as described in~\citet{Locci+2019}.

As discussed in SM21, the two outer planets, b and e, show radii that are relatively small with respect to the radii of planets of similar mass, suggesting the presence of a very massive core and/or 
an extreme enrichment in heavy elements that possibly occurred during their migratory phase. For this reason, we preferred to use for all planets the formulation of \citet{LopFor14} for the case of enhanced atmospheric opacity.

For each time step of the simulation, we updated the planetary mass (and consequently $f_{\rm atm}$) in response to the mass loss, and we employed again the relation by \citet{LopFor14} to compute a new value of $R_{\rm env}$. The latter quantity, added to the core radius, provided the updated planetary radius. 


% \bibliography{v1298}{}
\bibliographystyle{aasjournal}
% \input{ms.bbl}



\end{document}